\documentclass[sigconf,screen]{acmart}
\pdfoutput=1 
\AtBeginDocument{%
  \providecommand\BibTeX{{%
    \normalfont B\kern-0.5em{\scshape i\kern-0.25em b}\kern-0.8em\TeX}}}

\setcopyright{rightsretained}
\acmPrice{15.00}
\acmDOI{10.1145/3617574.3617858}
\acmYear{2023}
\copyrightyear{2023}
\acmSubmissionID{fsews23sesafemlmain-id4515-p}
\acmISBN{979-8-4007-0379-9/23/12}
\acmConference[SE4SafeML '23]{Proceedings of the 1st International Workshop on Dependability and Trustworthiness of Safety-Critical Systems with Machine Learned Components}{December 4, 2023}{San Francisco, CA, USA}
\acmBooktitle{Proceedings of the 1st International Workshop on Dependability and Trustworthiness of Safety-Critical Systems with Machine Learned Components (SE4SafeML '23), December 4, 2023, San Francisco, CA, USA}
\received{2023-07-04}
\received[accepted]{2023-08-10}

\usepackage{microtype}
\usepackage{tcolorbox}
\usepackage{dirtytalk}
\usepackage{comment}
\usepackage{booktabs}
\usepackage{array}
\usepackage{amsmath,amsfonts}
\usepackage{graphicx}
\usepackage{textcomp}
\usepackage{xcolor}
\usepackage{float}
\usepackage{hyperref}
\usepackage{subcaption}
\usepackage[normalem]{ulem}
\usepackage[linesnumbered,ruled,vlined]{algorithm2e}

\SetCommentSty{mycommfont}

\usepackage[utf8]{inputenc}
\usepackage{pgfplots}
\DeclareUnicodeCharacter{2212}{−}
\usepgfplotslibrary{groupplots,dateplot}
\usetikzlibrary{patterns,shapes.arrows}
\pgfplotsset{compat=newest}

\usepackage{fancyhdr}

\usepackage{xspace}

\usepackage{pifont}
\usepackage{pgfplots}
\usepackage{subcaption}
\usepackage{array}
\newcolumntype{L}[1]{>{\raggedright\let\newline\\\arraybackslash\hspace{0pt}}m{#1}}
\newcolumntype{C}[1]{>{\centering\let\newline\\\arraybackslash\hspace{0pt}}m{#1}}
\newcolumntype{R}[1]{>{\raggedleft\let\newline\\\arraybackslash\hspace{0pt}}m{#1}}

\usepackage[all]{nowidow}

\usetikzlibrary{patterns}
\usetikzlibrary{intersections}
\usetikzlibrary{calc}
\usetikzlibrary{arrows.meta}
\usetikzlibrary{shapes.misc}

\newcommand{\ie}{\emph{i.e.,}\xspace}
\newcommand{\eg}{\emph{e.g.,}\xspace}

\newcommand{\tool}{\textsc{FedDefender}\xspace}

\begin{document}

\title{FedDefender: Backdoor Attack Defense in Federated Learning}

\author{Waris Gill}
\affiliation{%
  \institution{Virginia Tech}
  \city{Blacksburg}
  \country{USA}}
\email{waris@vt.edu}

\author{Ali Anwar}
\affiliation{%
  \institution{University of Minnesota}
  \city{Minneapolis}
  \country{USA}}
\email{aanwar@umn.edu}

\author{Muhammad Ali Gulzar}
\affiliation{%
  \institution{Virginia Tech}
  \city{Blacksburg}
  \country{USA}}
\email{gulzar@cs.vt.edu}

\begin{abstract}
    Federated Learning (FL) is a privacy-preserving distributed machine learning technique that enables individual clients (\eg user participants, edge devices, or organizations) to train a model on their local data in a secure environment and then share the trained model with an aggregator to build a global model collaboratively. In this work, we propose \tool, a defense mechanism against targeted poisoning attacks in FL by leveraging differential testing. \tool first applies differential testing on  clients' models using a synthetic input. Instead of comparing the output (predicted label), which is unavailable for synthetic input, \tool fingerprints the neuron activations of clients' models to identify a potentially malicious client containing a backdoor. We evaluate \tool using MNIST and FashionMNIST datasets with 20 and 30 clients, and our results demonstrate that \tool effectively mitigates such attacks, reducing the attack success rate (ASR) to 10\% without deteriorating the global model performance.
\end{abstract}





\maketitle

\section{Introduction}
Federated Learning (FL) trains a global model on decentralized data from multiple clients without directly accessing their individual data samples. 
FL improves model accuracy by leveraging the combined data from multiple clients and also improves privacy by keeping individual data samples on the clients' devices.
However, the decentralized nature of FL makes it vulnerable to targeted poisoning attacks (often called backdoor attacks). In such attacks, an adversarial client manipulates its training data to inject a backdoor into a global model.  Since the server does not have access to the raw training data of the clients, such attacks remain hidden until a trigger is injected into the input. Therefore, it is highly challenging to detect and defend against backdoor attacks in FL \cite{bagdasaryan2020backdoor, ozdayi2021defending, wang2022towards, jebreel2022fl}.

Prior work \cite{sun2019can} on defending against targeted poisoning attacks in FL has focused on using norm clipping ({\em NormClipping}) to detect and mitigate these attacks. Norm clipping involves computing the norms of model updates received from clients and rejecting updates that exceed a certain threshold. This technique has been shown to be effective in some cases, but it has limitations. For example, if an attacker carefully crafts the attack such that the norm of the gradient is not noticeably large, the norm clipping approach will not be effective in detecting the attack. Therefore, alternative approaches are needed to defend against targeted poisoning attacks.

\noindent{\textit{\textbf{Contribution and Key Insight.}}} In this work, we propose \tool, a defense against backdoor attacks in federated learning by leveraging differential testing for FL \cite{gill2023feddebug}. \tool~minimizes the impact of a malicious client on the global model by limiting its contribution to the aggregated global model. Instead of comparing the predicted label of an input, which is often unavailable in FL, \tool fingerprints the neuron activations of clients' models on the same input and uses differential testing to identify potential malicious clients. Our insight is that since clients in FL have homogeneous models trained on similar concepts, their neuron activations should have some similarities on a given input \cite{gill2023feddebug}. At the central server, if a client's model displays neuron activation patterns that significantly differ from other clients (\ie majority of clients), such a client's model may contain a trigger pattern and can be flagged as potentially malicious.


\noindent{\textit{\textbf{Evaluations.}}} We evaluate \tool with 20 and 30 FL clients on MNIST and FashionMNIST datasets. Our results demonstrate that compared to the norm clipping defense \cite{sun2019can}, \tool~effectively defends against backdoor attacks and reduces the attack success rate (ASR) to 10\% without negatively impacting the global model accuracy. \tool's artifact is available at \url{https://github.com/warisgill/FedDefender}.

\section{Background and Related Work}
\label{background}

\noindent{\textit{\textbf{Federated Learning.}}} In Federated Learning (FL), multiple {\em clients} (\eg mobile devices, smart home devices, and autonomous vehicles) locally train models on their private training data. The trained client's model is sent back to a {\em central server} (also called an {\em aggregator}). A client's model comprises a collection of {\em weights} connecting {\em neurons} in a neural network. All client models work on structurally (same number of neurons and layer) same neural network. After the participating clients' models are received, the aggregator uses a {\em fusion} algorithm to merge all models into a single {\em global model}.  A {\em round} in FL starts with the client's training and ends once a global model is constructed. 
    Federated Averaging (FedAvg)~\cite{mcmahan2017communication} is a popular fusion algorithm that uses the following equation to build a global model using the client's models at each round.
    
    \vspace{-2ex}
    \begin{equation} \label{eq:1}
         W_{global}^{t+1} =  \sum_{k=1}^K \frac{n_k}{n} W_k^{(t)}  
    \end{equation}

    \vspace{-2ex}
    
    \noindent $W_k^{(t)}$ and $n_k$ represent weights and size of training data of client $k$ in a given round $t$, respectively. 
    The variable $n$ represents the total number of data points from all clients, and it is calculated as $n = \sum_{k=1}^{K} n_k$. 
 At the end of the round, the global model is sent back to all participating clients to be used as a {\em pretrained} model in their local training during the next round. A {\em malicious client} sends its incorrect model after injecting a backdoor during its local training to manipulate the global model.

    \noindent{\textit{\textbf{Differential Testing.}}} Differential testing is a software testing technique. It executes two or more comparable programs on the same test input and compares resulting outputs to identify unexpected behavior \cite{gulzar2019perception}. In prior work, it is used to find bugs in compilers \cite{yang2011finding}, deep neural networks \cite{pei2017deepxplore}, and faulty clients in FL\cite{gill2023feddebug}. 
    
    \noindent{\textit{\textbf{Backdoor Attack and Defense.}}} Backdoor attacks in the context of computer vision refer to a specific type of malicious behavior in which an attacker injects a \say{backdoor} into a machine learning (ML) model during its training \cite{wang2019neural}. This backdoor allows the attacker to gain control over the model by providing a specific input that triggers the model to behave in a way that is beneficial to the attacker. This type of attack is particularly concerning as models are often used for tasks such as object recognition, and the ability to manipulate these models can have significant real-world consequences. 
    For example, an attacker could train a model to recognize a stop sign but also include a hidden trigger that causes misclassification, leading to unsafe situations in the real world.

    In FL, a malicious client $k$ can inject a backdoor to the global model ($W^{t+1}$) by manipulating its local model $W_k^{(t)}$ \cite{sun2019can,ozdayi2021defending, wang2022towards, jebreel2022fl, bagdasaryan2020backdoor}. Prior approaches \cite{ozdayi2021defending, wang2022towards} propose defenses by changing the underlying FL training protocol (\eg changes in the FedAvg protocol). Such defenses require special alterations to work with other FL training protocols such as FedProx~\cite{li2020federated} and FedAvg~\cite{mcmahan2017communication}. Sun et al. \cite{sun2019can} propose norm clipping to detect and mitigate these attacks. Norm clipping can degrade the performance of a global model, and it can be easily bypassed with carefully crafted attacks.     
    Therefore, alternative approaches are needed that can be integrated with any fusing algorithm (\eg FedAvg~\cite{mcmahan2017communication}, FedProx~\cite{li2020federated}) without requiring any changes to fusion protocols and, at the same time, do not impact the performance of the global model while still protecting against backdoor attacks. 

\section{Threat Model}

    We consider a single malicious client (\ie attacker) participating in each round ($t$). The malicious client $k$ injects a square trigger pattern ($4$ x $4$) to its $n_k$ training images to manipulate its local model ($W_k^{(t)}$) during local training. The attacker can increase the strength of a backdoor attack $X$ times by scaling up its number of training data points $n_k$ (\eg $n_k \leftarrow n_k \cdot 20$) to successfully inject the backdoor into the global model ($W^{t+1}$) during aggregation (Equation~\ref{eq:1}). The goal of the attacker in this threat model is to gain control over the federated learning model by injecting a backdoor trigger and using it to manipulate the model's behavior. 

\section{FedDefender Design}
\label{section:defense}

    \begin{algorithm}[!ht]
    
        {\scriptsize
         \SetKwInput{KwInput}{Input}                
         \SetKwInput{KwOutput}{Output}              
        \DontPrintSemicolon
        \KwInput{Let $client2weights = \{w_1, w_2, ... w_k\}$ be a dictionary that maps k clients to their models' parameters }
        \KwInput{Let $N =  \{n_1, n_2, ... n_k\}$ be number of training examples of k clients}
        \KwInput{Let $test\_inputs$ be a list of inputs}

        \KwInput{Let $\theta$ be a threshold for malicious confidence}


        \KwOutput{$W^{t+1}$: global model for the next round}
        
        \SetKwProg{Fn}{Function}{:}{}
        {
            {
                $client2mal\_confidence = Dictionary()$ \tcp{ An empty dictionary}
                $min\_n_k  = min(N)$  \tcp{Get a minimum number of training examples among k clients}
                \For{each $input\_i \in test\_inputs$}{
                    \tcp {Find the potential malicious client using FL differential testing technique \cite{gill2023feddebug}}
                    
                    $potential\_mal\_client =  FL\_DifferentialTesting(client2weights, input\_i, 0)$ \;
                    \tcp {Increment the confidence of a potential malicious client}
                    $client2mal\_confidence[potential\_mal\_client] += 1$ \; 
                }
                
                \tcp {Defense by restricting potential malicious clients contribution}
                \For {each $client \in client2mal\_confidence$}{
                    \tcp {Normalize the confidence of each client}
                    $client2mal\_confidence[client] = client2mal\_confidence[client]/ len(test\_inputs)$
                
                    \tcp {If threshold satisfied, discard malicious client's contribution by setting their number of training examples to 0}
                    \If { $client2mal\_confidence[client] > \theta$} {
                    $    N[client] = 0$
                    }
                    \Else{
                        
                        $N[client] =  int(min\_n_k * (1 - client2mal\_confidence[client]))$ \tcp {
                        Reduce training examples if the client seems less malicious to penalize their contribution to the global model.}
                        }
                }                
                
                $W^{t+1} = FedAvg(client2weights, N)$ \tcp {Compute the global model using FedAvg}
                \KwRet $W^{t+1}$ \;
            }}
          }
        \caption {\tool  Defense}
        \label{algo:defense}

        \end{algorithm}

    To achieve optimal performance of the global model and protect its integrity, it is critical to correctly identify the potential malicious clients and restrict their participation in the global model $W^{t+1}$ before the aggregation step (Equation~\ref{eq:1}). Access to clients' data is prohibited in FL, and collecting new test data at the central server has its own challenges. Such challenges make existing backdoor detection techniques \cite{wang2019neural} impractical. Thus, backdoor detection in FL requires a novel solution to mitigate the backdoor attack without any dependence on real-world test data. 

    \noindent{\textit{\textbf{Differential Testing FL Clients.}}} Gill et al. \cite{gill2023feddebug} propose a differential technique to find faulty clients in an FL round training without requiring access to real-world test inputs. It generates inputs randomly at the central server and compares the behaviors of clients' models at the neuron level to localize a faulty client. The internal neuron values of the models are used as a fingerprint of the behavior on the given input, and a client is flagged as malicious if its behavior deviates significantly from the majority of the clients. The key insight is that the behavior of a malicious client's model will be different from that of benign clients, as malicious executions are inherently different from correct ones. We use a neuron activation threshold equal to zero to profile the behavior (\ie neuron activations) of a client model. Note techniques such as GradCAM \cite{selvaraju2017grad}, DeepLift \cite{shrikumar2017learning, ancona2017towards}, DeepLiftShap \cite{lundberg2017unified}, or Internal Influence \cite{leino2018influence} can also be used to profile neuron activations.  

    \tool adapts differential testing technique for FL \cite{gill2023feddebug} to detect behavioral discrepancies among clients' models, with the aim of identifying potential malicious clients in a given FL training round. Algorithm~\ref{algo:defense} outlines the defense strategy of \tool against backdoor attacks in FL. The inputs to Algorithm~\ref{algo:defense} include the clients' models ($client2weights$), a list containing the number of training examples for each client ($N$), a set of randomly generated test inputs ($test\_inputs$), and a threshold for malicious confidence $\theta$. \tool first employs the differential execution technique, as outlined in \cite{gill2023feddebug}, to identify a potential malicious client on each input. It then updates the corresponding client's malicious score (lines 3-5).  Subsequently, \tool limits the contribution of a client if its malicious confidence exceeds the specified threshold $\theta$ (lines 6-11). Finally, the global model is computed using the updated contribution of clients (line 12). As an illustration, consider a scenario in which ten clients are participating in a given FL training round, the malicious threshold is set at 0.5, and 100 test inputs are generated. \tool computes the malicious confidence of all clients. Clients 1, 3, and 7 have malicious confidence scores of 20/100, 60/100, and 20/100, respectively. The remaining clients have a malicious confidence score of zero. \tool discards the contribution of client 3 as it exceeds the malicious threshold and accordingly limits the contributions of the other clients.
    

    \begin{figure}[t]
        \centering
        \resizebox{0.5\textwidth}{!}
        {

\begin{tikzpicture}
  
    \begin{groupplot}[
        group style={group size=3 by 2, horizontal sep= 2cm, vertical sep= 3cm},
        xmin=0,
        ymin=0,
        xlabel=Training Round,
        tick label style={font=\huge},
        label style={font=\huge},
        title style={font=\huge},
        xmin = 0.1,
        tick align=outside, 
        tick pos=both,
      ]
      
      \nextgroupplot[title=Attack Scale: 1X, ylabel= Accuracy (\%), ylabel style={align=center}]
      \addplot [line width=2.4pt, black, smooth,tension=0.15, mark=square, color=red]  table [x=x, y=y_1X_ASR, col sep=comma] {attack_scale.csv};
      \addplot [line width=2.4pt, black, densely dotted, smooth,tension=0.15, mark=triangle*, color=green]  table [x=x, y=y_1X_CA, col sep=comma] {attack_scale.csv};
      
      \nextgroupplot[title=Attack Scale: 3X, legend style={at={(0.5,1.35)}, anchor=north, legend columns=2, font=\huge},legend entries={Attack Success Rate (ASR), Model Classification Accuracy}, ]
      \addplot [line width=2.4pt, black, smooth,tension=0.15, mark=square, color=red]  table [x=x, y=y_3X_ASR, col sep=comma] {attack_scale.csv};
      \addplot [line width=2.4pt, black, densely dotted, smooth,tension=0.15, mark=triangle*, color=green]  table [x=x, y=y_3X_CA, col sep=comma] {attack_scale.csv};

      \nextgroupplot[title=Attack Scale: 5X, ylabel style={align=center}]
      \addplot [line width=2.4pt, black, smooth,tension=0.15, mark=square, color=red]  table [x=x, y=y_5X_ASR, col sep=comma] {attack_scale.csv};
      \addplot [line width=2.4pt, black, densely dotted, smooth,tension=0.15, mark=triangle*, color=green]  table [x=x, y=y_5X_CA, col sep=comma] {attack_scale.csv};

      \nextgroupplot[title=Attack Scale: 10X, ylabel= Accuracy (\%), ylabel style={align=center}]
      \addplot [line width=2.4pt, black, smooth,tension=0.15, mark=square, color=red]  table [x=x, y=y_10X_ASR, col sep=comma] {attack_scale.csv};
      \addplot [line width=2.4pt, black, densely dotted, smooth,tension=0.15, mark=triangle*, color=green]  table [x=x, y=y_10X_CA, col sep=comma] {attack_scale.csv};

      \nextgroupplot[title=Attack Scale: 20X, ylabel style={align=center}]
      \addplot [line width=2.4pt, black, smooth,tension=0.15, mark=square, color=red]  table [x=x, y=y_20X_ASR, col sep=comma] {attack_scale.csv};
      \addplot [line width=2.4pt, black, densely dotted, smooth,tension=0.15, mark=triangle*, color=green]  table [x=x, y=y_20X_CA, col sep=comma] {attack_scale.csv};
    \end{groupplot}
  \end{tikzpicture}}
        \caption{The scaling factor increases the strength of the malicious client by increasing the number of training examples, $n_k$, by a factor of $X$.  This enhances the chances of successfully injecting a backdoor in the global model $W^{t+1}$. }
        \label{fig:attack_scale}
        
    \end{figure}
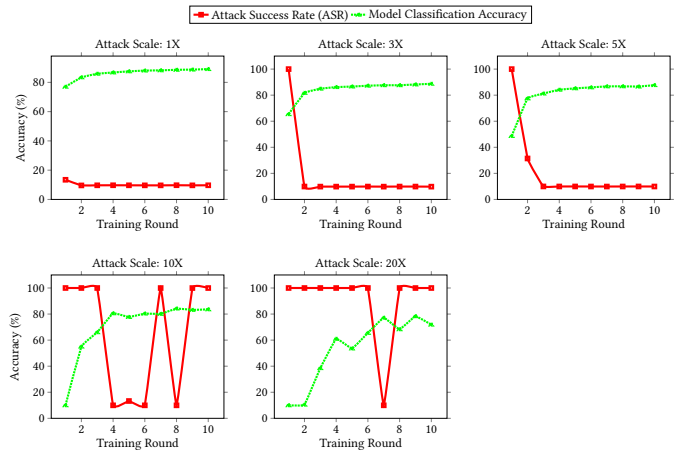

    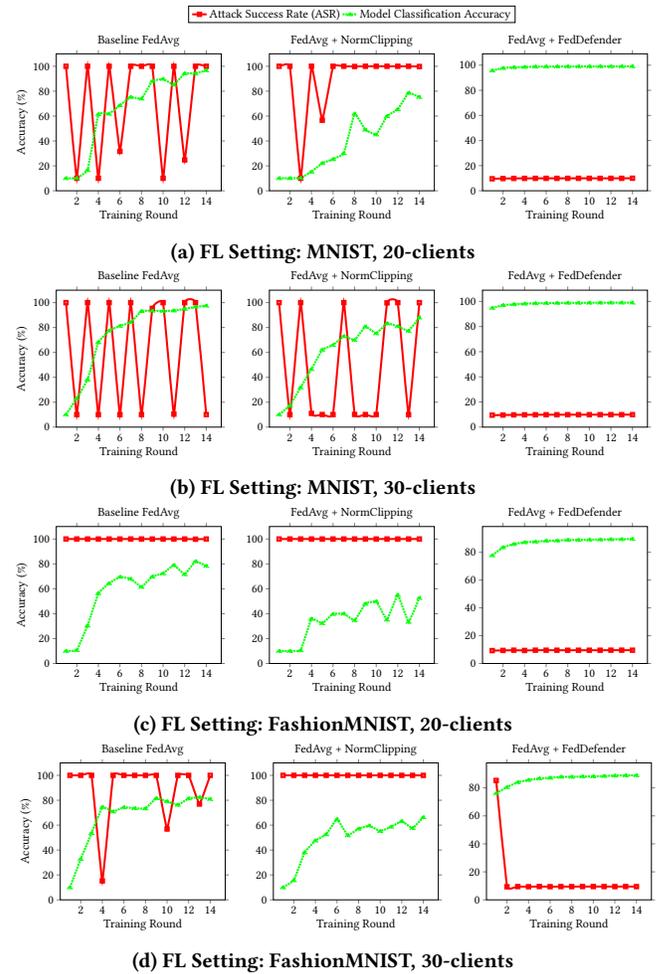
\begin{figure}[t]
        \begin{subfigure}{.32\textwidth}
            \hspace{-0.5in}\scalebox{0.32}
            {\begin{tikzpicture}  
    \begin{groupplot}[
        group style={group size=3 by 1, horizontal sep= 2cm,},
        xmin=0,
        ymin=0,
        xlabel=Training Round,
        tick label style={font=\huge},
        label style={font=\huge},
        title style={font=\huge},
        xmin = 0.1,
        tick align=outside, 
        tick pos=both,
      ]
      
      \nextgroupplot[title=Baseline FedAvg , ylabel= Accuracy (\%), ylabel style={align=center}]
      \addplot [line width=2.4pt, black, smooth,tension=0.15, mark=square, color=red]  table [x=x, y=y_FedAvg_ASR, col sep=comma] {defense1-dataset-mnist,-totalclients-20,-total_rounds-14,.csv};
      \addplot [line width=2.4pt, black, densely dotted, smooth,tension=0.15, mark=triangle*, color=green]  table [x=x, y=y_FedAvg_CA, col sep=comma] {defense1-dataset-mnist,-totalclients-20,-total_rounds-14,.csv};
      
      \nextgroupplot[title=FedAvg + NormClipping, legend style={at={(0.5,1.35)}, anchor=north, legend columns=2, font=\huge},legend entries={Attack Success Rate (ASR), Model Classification Accuracy}, ]
      \addplot [line width=2.4pt, black, smooth,tension=0.15, mark=square, color=red]  table [x=x, y=y_NormClipping_ASR, col sep=comma] {defense1-dataset-mnist,-totalclients-20,-total_rounds-14,.csv};
      \addplot [line width=2.4pt, black, densely dotted, smooth,tension=0.15, mark=triangle*, color=green]  table [x=x, y=y_NormClipping_CA, col sep=comma] {defense1-dataset-mnist,-totalclients-20,-total_rounds-14,.csv};

      \nextgroupplot[title=FedAvg + FedDefender, ylabel style={align=center}]
      \addplot [line width=2.4pt, black, smooth,tension=0.15, mark=square, color=red]  table [x=x, y=y_FedDefender_ASR, col sep=comma] {defense1-dataset-mnist,-totalclients-20,-total_rounds-14,.csv};
      \addplot [line width=2.4pt, black, densely dotted, smooth,tension=0.15, mark=triangle*, color=green]  table [x=x, y=y_FedDefender_CA, col sep=comma] {defense1-dataset-mnist,-totalclients-20,-total_rounds-14,.csv};

    \end{groupplot}
  \end{tikzpicture}
  }
            \caption{FL Setting: MNIST, 20-clients}
            \label{fig:defense-1}
        \end{subfigure}
    
        \begin{subfigure}{.32\textwidth}
            \hspace{-0.5in}\scalebox{0.32}
            {\begin{tikzpicture}  
    \begin{groupplot}[
        group style={group size=3 by 1, horizontal sep= 2cm,},
        xmin=0,
        ymin=0,
        xlabel=Training Round,
        tick label style={font=\huge},
        label style={font=\huge},
        title style={font=\huge},
        xmin = 0.1,
        tick align=outside, 
        tick pos=both,
      ]
      
      \nextgroupplot[title=Baseline FedAvg , ylabel= Accuracy (\%), ylabel style={align=center}]
      \addplot [line width=2.4pt, black, smooth,tension=0.15, mark=square, color=red]  table [x=x, y=y_FedAvg_ASR, col sep=comma] {defense1-dataset-mnist,-totalclients-30,-total_rounds-14,.csv};
      \addplot [line width=2.4pt, black, densely dotted, smooth,tension=0.15, mark=triangle*, color=green]  table [x=x, y=y_FedAvg_CA, col sep=comma] {defense1-dataset-mnist,-totalclients-30,-total_rounds-14,.csv};
      
      \nextgroupplot[title=FedAvg + NormClipping,]
      \addplot [line width=2.4pt, black, smooth,tension=0.15, mark=square, color=red]  table [x=x, y=y_NormClipping_ASR, col sep=comma] {defense1-dataset-mnist,-totalclients-30,-total_rounds-14,.csv};
      \addplot [line width=2.4pt, black, densely dotted, smooth,tension=0.15, mark=triangle*, color=green]  table [x=x, y=y_NormClipping_CA, col sep=comma] {defense1-dataset-mnist,-totalclients-30,-total_rounds-14,.csv};

      \nextgroupplot[title=FedAvg + FedDefender, ylabel style={align=center}]
      \addplot [line width=2.4pt, black, smooth,tension=0.15, mark=square, color=red]  table [x=x, y=y_FedDefender_ASR, col sep=comma] {defense1-dataset-mnist,-totalclients-30,-total_rounds-14,.csv};
      \addplot [line width=2.4pt, black, densely dotted, smooth,tension=0.15, mark=triangle*, color=green]  table [x=x, y=y_FedDefender_CA, col sep=comma] {defense1-dataset-mnist,-totalclients-30,-total_rounds-14,.csv};

    \end{groupplot}
  \end{tikzpicture}
  }
            \caption{FL Setting: MNIST, 30-clients}
            \label{fig:defense-2}
        \end{subfigure}
    
        \begin{subfigure}{0.32\textwidth}
            \hspace{-0.5in}\scalebox{0.32}
            {\begin{tikzpicture}  
    \begin{groupplot}[
        group style={group size=3 by 1, horizontal sep= 2cm,},
        xmin=0,
        ymin=0,
        xlabel=Training Round,
        tick label style={font=\huge},
        label style={font=\huge},
        title style={font=\huge},
        xmin = 0.1,
        tick align=outside, 
        tick pos=both,
      ]
      
      \nextgroupplot[title=Baseline FedAvg , ylabel= Accuracy (\%), ylabel style={align=center}]
      \addplot [line width=2.4pt, black, smooth,tension=0.15, mark=square, color=red]  table [x=x, y=y_FedAvg_ASR, col sep=comma] {defense1-dataset-fashionmnist,-totalclients-20,-total_rounds-14,.csv};
      \addplot [line width=2.4pt, black, densely dotted, smooth,tension=0.15, mark=triangle*, color=green]  table [x=x, y=y_FedAvg_CA, col sep=comma] {defense1-dataset-fashionmnist,-totalclients-20,-total_rounds-14,.csv};
      
      \nextgroupplot[title=FedAvg + NormClipping,  ]
      \addplot [line width=2.4pt, black, smooth,tension=0.15, mark=square, color=red]  table [x=x, y=y_NormClipping_ASR, col sep=comma] {defense1-dataset-fashionmnist,-totalclients-20,-total_rounds-14,.csv};
      \addplot [line width=2.4pt, black, densely dotted, smooth,tension=0.15, mark=triangle*, color=green]  table [x=x, y=y_NormClipping_CA, col sep=comma] {defense1-dataset-fashionmnist,-totalclients-20,-total_rounds-14,.csv};

      \nextgroupplot[title=FedAvg + FedDefender, ylabel style={align=center}]
      \addplot [line width=2.4pt, black, smooth,tension=0.15, mark=square, color=red]  table [x=x, y=y_FedDefender_ASR, col sep=comma] {defense1-dataset-fashionmnist,-totalclients-20,-total_rounds-14,.csv};
      \addplot [line width=2.4pt, black, densely dotted, smooth,tension=0.15, mark=triangle*, color=green]  table [x=x, y=y_FedDefender_CA, col sep=comma] {defense1-dataset-fashionmnist,-totalclients-20,-total_rounds-14,.csv};

    \end{groupplot}
  \end{tikzpicture}
  }
            \caption{FL Setting: FashionMNIST, 20-clients}
            \label{fig:defense-3}
        \end{subfigure}
    
        \begin{subfigure}{0.32\textwidth}
           \hspace{-0.5in} \scalebox{0.32}
            {\begin{tikzpicture}  
    \begin{groupplot}[
        group style={group size=3 by 1, horizontal sep= 2cm,},
        xmin=0,
        ymin=0,
        xlabel=Training Round,
        tick label style={font=\huge},
        label style={font=\huge},
        title style={font=\huge},
        xmin = 0.1,
        tick align=outside, 
        tick pos=both,
      ]
      
      \nextgroupplot[title=Baseline FedAvg , ylabel= Accuracy (\%), ylabel style={align=center}]
      \addplot [line width=2.4pt, black, smooth,tension=0.15, mark=square, color=red]  table [x=x, y=y_FedAvg_ASR, col sep=comma] {defense1-dataset-fashionmnist,-totalclients-30,-total_rounds-14,.csv};
      \addplot [line width=2.4pt, black, densely dotted, smooth,tension=0.15, mark=triangle*, color=green]  table [x=x, y=y_FedAvg_CA, col sep=comma] {defense1-dataset-fashionmnist,-totalclients-30,-total_rounds-14,.csv};
      
      \nextgroupplot[title=FedAvg + NormClipping, ]
      \addplot [line width=2.4pt, black, smooth,tension=0.15, mark=square, color=red]  table [x=x, y=y_NormClipping_ASR, col sep=comma] {defense1-dataset-fashionmnist,-totalclients-30,-total_rounds-14,.csv};
      \addplot [line width=2.4pt, black, densely dotted, smooth,tension=0.15, mark=triangle*, color=green]  table [x=x, y=y_NormClipping_CA, col sep=comma] {defense1-dataset-fashionmnist,-totalclients-30,-total_rounds-14,.csv};

      \nextgroupplot[title=FedAvg + FedDefender, ylabel style={align=center}]
      \addplot [line width=2.4pt, black, smooth,tension=0.15, mark=square, color=red]  table [x=x, y=y_FedDefender_ASR, col sep=comma] {defense1-dataset-fashionmnist,-totalclients-30,-total_rounds-14,.csv};
      \addplot [line width=2.4pt, black, densely dotted, smooth,tension=0.15, mark=triangle*, color=green]  table [x=x, y=y_FedDefender_CA, col sep=comma] {defense1-dataset-fashionmnist,-totalclients-30,-total_rounds-14,.csv};

    \end{groupplot}
  \end{tikzpicture}
  }
            \caption{FL Setting: FashionMNIST, 30-clients}
            \label{fig:defense-4}
        \end{subfigure}
    
        \caption{Comparison of \tool, with the baseline FedAvg and NormClipping defense mechanisms. Figures indicate that \tool successfully mitigates the attack and lowers the ASR close to 10\% .}
        \label{fig:defense-1-4}
        
    \end{figure}

\section{Evaluation}

    We evaluate \tool on (1) Attack Success Rate (ASR) \cite{wang2019neural} and (2) classification performance of the global model. 
    
    \noindent{\textit{\textbf{Dataset, Model, FL Framework.}} We use MNIST~\cite{cirecsan2011high} and FashionMNIST~\cite{xiao2017/online} datasets. Each dataset contains 60K training and 10K testing grayscale, 28x28 images spanning ten different classes. The data is randomly distributed without any overlapping data points among FL clients. Each client trains a convolutional neural network (CNN). The CNN architecture is outlined in ~\cite{Training65:online}. We set the learning rate to 0.001, epochs equal to 5 and 15, batch size of 32, and trained each configuration for at least 10 rounds. We implement our approach in Flower FL framework~\cite{beutel2020flower}. We run our experiments on AMD 16-core processor with 128 GB RAM.

    \noindent{\textit{\textbf{Evaluation Metrics.}} We used the attack success rate (ASR) \cite{wang2019neural} and classification accuracy of the global model to compare \tool with norm clipping defense \cite{sun2019can}.


\noindent{\textit{\textbf{Backdoor Attack Strength.}} The strength of a backdoor attack (on the global model $W^{t+1}$) can be evaluated by considering the injection of a 4x4 trigger pattern into the training data of a malicious client, as well as the scaling of the number of examples used for such injection. Figure~\ref{fig:attack_scale} demonstrates the effect of varying attack scales on the attack success rate (ASR) in an FL configuration consisting of 20 clients with the FashionMNIST dataset. Without scaling, \ie $Attack\ Scale = 1\times$, a malicious client is unable to inject a backdoor into the global model successfully. For the remaining experiments, a 20$\times$ scale is used to represent the maximum strength of the backdoor attack.

\noindent{\textit{\textbf{Backdoor Defense Evaluation.}}} We compare \tool with the baseline Federated Averaging (FedAvg) algorithm (\ie without any defense)~\cite{mcmahan2017communication}  and the $NormClipping$ defense mechanism~\cite{sun2019can}, using 20 and 30 FL clients configurations. The MNIST and FashionMNIST datasets are used in these experiments. Each setting is trained for 14 rounds, with 5 epochs in each round. The results of these experiments are illustrated in Figure~\ref{fig:defense-1-4}, with the x-axis representing the number of training rounds and the y-axis representing the accuracy. The attack success rate (ASR) and classification accuracy are used to compare \tool with the baseline and $NormClipping$ defense mechanisms. A lower ASR indicates that the malicious client is unable to manipulate the global model behavior using its backdoor. As shown in Figures \ref{fig:defense-1}-\ref{fig:defense-4}, the $NormClipping$ defense fails to provide any defense against the backdoor attack and also negatively impacts the global model's ($W^{t+1}$) classification accuracy. In contrast, \tool successfully mitigates the attack and lowers the ASR close to 10\% without deteriorating the global model's classification accuracy.


\noindent{\textit{\textbf{Malicious Confidence Threshold ($\theta$).}}}  The impact of the malicious confidence threshold ($\theta$) in Algorithm~\ref{algo:defense} on the mitigation of the backdoor attack is also examined. Figure~\ref{fig:theat_impact} shows the results of this analysis, using an FL configuration of 20 clients trained on the MNIST dataset. Each client model is trained for 15 epochs. 
Figure~\ref{fig:theat_impact} illustrates that unless the potential malicious client is penalized aggressively, \tool is incapable of mitigating the attack. To aggressively penalize a client, the client's contribution is ignored before aggregation (lines 8-11 of Algorithm~\ref{algo:defense}).


    \begin{tcolorbox}[left=0mm, right=0mm, top=0mm, bottom=0mm]
        \textbf{Takeaway:} \tool successfully protects against backdoor attacks without impacting the global model classification accuracy.
    \end{tcolorbox}

\noindent{\textit{\textbf{\tool False Positive Rate.}}} We evaluate the false positive rate to assess the impact of \tool on the global model's classification accuracy using a federated learning (FL) setting of 20 clients and the FashionMNIST dataset. In this scenario, all clients are benign, that is, there is no malicious client present. As shown in Figure~\ref{fig:false-positives}, \tool hardly produces any false positives and demonstrates similar performance as the baseline Federated Averaging (FedAvg) and $NormClipping$ defense mechanisms.

    \begin{tcolorbox}[left=0mm, right=0mm, top=0mm, bottom=0mm]
    \textbf{Takeaway:} \tool does not impact the global model accuracy, even if there is no malicious client.
    \end{tcolorbox}

\noindent{\textit{\textbf{Threat to Validity.}}} To address potential threats to external validity, we perform experiments on two standardized FL datasets. Additionally, to mitigate potential threats arising from randomness in the \tool's random input generation, we evaluate each configuration on at least 100 random test inputs to compute the malicious confidence of a client. Despite these measures, certain threats to the validity of the experiments, such as variations in data distribution across clients, neuron activation threshold (default is zero), size of random test input, and type of convolutional neural networks (CNNs) may still exist. Future research will explore these potential threats in greater detail.

  \begin{figure}[!ht]
        \resizebox{0.5\textwidth}{!}
        {\begin{tikzpicture}  
    \begin{groupplot}[
        group style={group size=3 by 1, horizontal sep= 2cm,},
        xmin=0,
        ymin=0,
        xlabel=Training Round,
        tick label style={font=\huge},
        label style={font=\huge},
        title style={font=\huge},
        xmin = 0.1,
        tick align=outside, 
        tick pos=both,
      ]
      
      \nextgroupplot[title=a) Baseline FedAvg , ylabel= Accuracy (\%), ylabel style={align=center}]
      \addplot [line width=2.4pt, black, smooth,tension=0.15, mark=square, color=red]  table [x=x, y=y_FedAvg_ASR, col sep=comma] {defense1-dataset-mnist,-totalclients-20,-total_rounds-15,.csv};
      \addplot [line width=2.4pt, black, densely dotted, smooth,tension=0.15, mark=triangle*, color=green]  table [x=x, y=y_FedAvg_CA, col sep=comma] {defense1-dataset-mnist,-totalclients-20,-total_rounds-15,.csv};
      
      \nextgroupplot[title=b) FedAvg + NormClipping, legend style={at={(0.5,1.35)}, anchor=north, legend columns=2, font=\huge},legend entries={Attack Success Rate (ASR), Model Classification Accuracy}, ]
      \addplot [line width=2.4pt, black, smooth,tension=0.15, mark=square, color=red]  table [x=x, y=y_NormClipping_ASR, col sep=comma] {defense1-dataset-mnist,-totalclients-20,-total_rounds-15,.csv};
      \addplot [line width=2.4pt, black, densely dotted, smooth,tension=0.15, mark=triangle*, color=green]  table [x=x, y=y_NormClipping_CA, col sep=comma] {defense1-dataset-mnist,-totalclients-20,-total_rounds-15,.csv};

      \nextgroupplot[title=c) FedAvg + FedDefender, ylabel style={align=center}]
      \addplot [line width=2.4pt, black, smooth,tension=0.15, mark=square, color=red]  table [x=x, y=y_FedDefender_ASR, col sep=comma] {defense1-dataset-mnist,-totalclients-20,-total_rounds-15,.csv};
      \addplot [line width=2.4pt, black, densely dotted, smooth,tension=0.15, mark=triangle*, color=green]  table [x=x, y=y_FedDefender_CA, col sep=comma] {defense1-dataset-mnist,-totalclients-20,-total_rounds-15,.csv};

    \end{groupplot}
  \end{tikzpicture}
  }
        \caption{Evaluation of the impact of the malicious confidence threshold. \tool is unable to mitigate the attack if the potential malicious client is not aggressively penalized.
        }
        \label{fig:theat_impact}
    \end{figure}
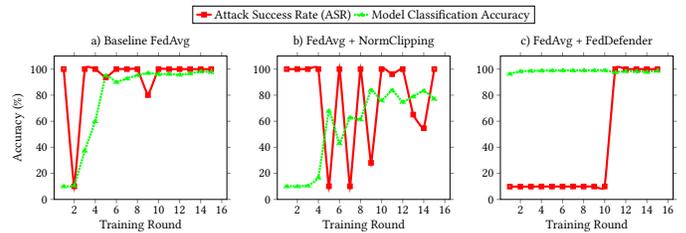

    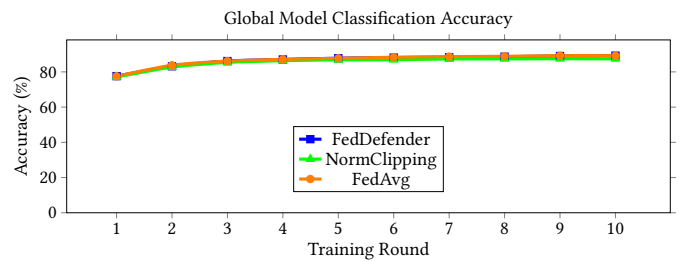
\begin{figure}[!ht]
        \centering
        \resizebox{0.5\textwidth}{!}
        {\begin{tikzpicture}  
    \begin{groupplot}[
        group style={group size=1 by 1},
        height=0.35\textwidth,
        width=1\textwidth,
        xmin=0,
        ymin=0,
        xlabel=Training Round,
        tick label style={font=\huge},
        label style={font=\huge},
        title style={font=\huge},
        xmin = 0.1,
        tick align=outside, 
        tick pos=both,
        legend style={at={(0.5,.5)}, anchor=north, legend columns=1, font=\huge},
        legend entries={FedDefender, NormClipping, FedAvg},
      ]
      
      \nextgroupplot[title=Global Model Classification Accuracy, ylabel= Accuracy (\%), ylabel style={align=center}]
      \addplot [line width=2.4pt, black, smooth, mark=square, color=blue]  table [x=x, y=FedDefender, col sep=comma] {benign_acc.csv};
      \addplot [line width=2.4pt, black, smooth, mark=triangle*, color=green]  table [x=x, y=NormClipping, col sep=comma] {benign_acc.csv};
      \addplot [line width=2.4pt, black, smooth,, mark=*, color=orange]  table [x=x, y=FedAvg, col sep=comma] {benign_acc.csv};

    \end{groupplot}
  \end{tikzpicture}
  }
        \caption{\tool performs similarly to FedAvg and NormClipping without penalizing benign clients. 
        }
        \label{fig:false-positives}
    \end{figure}

\section{Future Work and Conclusion}

\noindent{\textit{\textbf{Future Work.}}} In future work, we propose to evaluate the potential of \tool by assessing its performance under various FL training settings. This could include varying the number of malicious clients, the number of training epochs, and data distribution across clients (\ie non-IID data distributions). Additionally, efforts could be made to further improve the detection capabilities of \tool, allowing precise identification of {\em multiple} malicious clients and reverse engineering their corresponding backdoor trigger patterns.

Another avenue of research would be to analyze the aggregation overhead of \tool compared to traditional aggregation protocols in FL. Extending the applicability of \tool to other model architectures, such as Transformers, which are commonly used in natural language processing tasks and speech recognition models, could be explored. Finally, incorporating realistic synthetic test inputs generated using generative adversarial networks (GANs) into the evaluation process could provide further insight into the performance of \tool.

\noindent{\textit{\textbf{Conclusion.}}} 
Our position is that traditional software testing principles have matured over the years and have provably improved the state of testing software; therefore, FL should benefit from similar advancements. In this work, we propose \tool, a defense mechanism against targeted poisoning attacks in FL that utilizes random test generation with differential testing. We demonstrate that \tool effectively detects and mitigates such attacks, reducing the ASR to 10\% without negatively impacting the global model accuracy. Our results show that \tool is more effective than the norm clipping defense and the baseline Federated Averaging (FedAvg) algorithm.


\bibliographystyle{ACM-Reference-Format}
\bibliography{main}

\end{document}